\documentclass[twocolumn,showpacs,preprintnumbers]{revtex4}

\def\AJ{{\it Ap. J.} }

\def\ASAS{{\it Astron. and Astrophys.} }

\def\CQG{{\it Class. Quantum Gravity} }

\def\GRG{{\it Gen. Relativity and Gravitation} }

\def\MPL{{\it Mod. Phys. Lett.} }
\def\MNRAS{{\it Mon. Not. R. Ast. Soc.} }
\def\NAT{{\it Nature} }

\def\PL{{\it Phys. Lett.} }
\def\PR{{\it Phys. Rev.} }
\def\PRL{{\it Phys. Rev. Lett.} }

\def\al{\alpha} \def\be{\beta} \def\ga{\gamma} \def\de{\delta}
\def\ep{\epsilon}   
   \def\ka{\kappa}
\def\la{\lambda}   
\def\si{\sigma}   
   
\def\om{\omega}   
\def\La{\Lambda}   
  \def\mn{{\mu\nu}}

 \def\frac#1#2{{\textstyle{{#1}\over
{#2}}}} 
\def\lsim{\mathrel{\rlap{\lower4pt\hbox{\hskip1pt$\sim$}}
\raise1pt\hbox{$<$}}}
\def\gsim{\mathrel{\rlap{\lower4pt\hbox{\hskip1pt$\sim$}}
\raise1pt\hbox{$>$}}} \def\sqr#1#2{{\vcenter{\vbox{\hrule height.#2pt
\hbox{\vrule width.#2pt height#1pt \kern#1pt \vrule width.#2pt} \hrule
height.#2pt}}}}
\def\square{\mathchoice\sqr66\sqr66\sqr{2.1}3\sqr{1.5}3}
\def\beq{\begin{equation}} \def\eeq{\end{equation}}
\def\beqa{\begin{eqnarray}} \def\eeqa{\end{eqnarray}}

\begin{document}

\title{On the non-minimal gravitational coupling to matter}

\vskip 0.2cm

\author{O. Bertolami}
\email{orfeu@cosmos.ist.utl.pt}

\author{J. P\'aramos}
\email{jorge.paramos@ist.utl.pt}

\vskip 0.2cm

\affiliation{Instituto Superior T\'ecnico, Departamento de
F\'{\i}sica\footnote{Also at Instituto de Plasmas e Fus\~ao Nuclear, Instituto Superior T\'ecnico.}, \\Av. Rovisco Pais 1, 1049-001 Lisboa, Portugal.}

\vskip 0.5cm

\date{\today}

\begin{abstract}

The connection between $f(R)$ theories of gravity and scalar-tensor models with a ``physical'' metric coupled to the scalar field is well known. In this work, we pursue the equivalence between a suitable scalar theory and a model that generalises the $f(R)$ scenario, encompassing both a non-minimal scalar curvature term and a non-minimum coupling of the scalar curvature and matter. This equivalence allows for the calculation of the PPN parameters $\beta$ and $\gamma$ and, eventually, a solution to the debate concerning the weak-field limit of $f(R)$ theories.

\vskip 0.5cm

\end{abstract}

\pacs{04.20.Fy, 04.80.Cc, 11.10.Ef \hspace{2cm}Preprint DF/IST-2.2008}

\maketitle


\section{Introduction}

Contemporary cosmology is faced with the outstanding challenge of understanding the existence and nature of the so-called dark components of the Universe: dark energy and dark matter. The former is required to explain the accelerated expansion of the Universe, and accounts for about $74 \% $ of the energy content of the Universe; the latter is hinted, for instance, by the flattening of galactic rotation curves and cluster dynamics \cite{bullet}, and constitutes about $22 \%$ of the Universe's energy budget. Several theories have been put forward to address these issues, usually resorting to the introduction of new fields; for dark energy, the so-called ``quintessence'' models consider the slow-roll down of a scalar field, thus inducing the observed accelerated expansion \cite{scalar, Amendola}. For dark matter, several weak-interacting particles (WIMPs) have been suggested, many arising from extensions to the Standard Model ({\it e.g.} axions, neutralinos). A scalar field can also account for an unified model of dark energy and dark matter \cite{Rosenfeld}. Alternatively, one can implement this unification through an exotic equation of state, such as the generalized Chaplygin gas \cite{Chaplygin}.

Other approaches consider that these observational challenges do not demand the inclusion of extra energy content in the Universe but, instead, they hint at an incompleteness of the fundamental laws and tenets of General Relativity (GR). Following this line of reasoning, one may resort {\it e.g.} to extensions of the Friedmann equation to include higher order terms in the energy density $\rho$ have been suggested (see {\it e.g.} \cite{Maartens} and references therein). Another approach considers changes on the fundamental action functional: a rather straight forward approach lies in replacing the linear scalar curvature term in the Einstein-Hilbert action by a function of the scalar curvature, $f(R)$; alternatively, one could resort to other scalar invariants of the theory \cite{f(R)} (see \cite{paramos} and references therein for a discussion).

As with several other theories \cite{solar1,solar2}, solar system tests could shed some light onto the possible form and behaviour of these $f(R)$ theories; amongst other considerations, this approach is based either in the more usual metric affine connection, or in the so-called Palatini approach \cite{Palatini}, where both the metric and the affine connection are taken as independent variables. As an example of a phenomenological consequence of this extension of GR, it has been shown that $f(R) \propto R^n$ theories yield a gravitational potential which displays an increasing, repulsive contribution, besides the Newtonian term \cite{flat}.

Another line of action lies in the comparison between present and future observational signatures and the parameterized post-Newtonian (PPN) metric coefficients arising from this extension of GR, taken in the weak field limit and when the added degree of freedom may be characterized by a light scalar field \cite{solar2} . Regarding this, some disagreement exists in the community, some arguing that no significant changes are predicted at a post-Newtonian level (see {\textit e.g.} \cite{PPN} and references therein); others defending that $f(R)$ theories yield the PPN parameter $\ga = 1/2$, which is clearly disallowed by the current experimental constraint $\ga - 1 = (2.1 \pm 2.3)\times 10^{-5}$ \cite{Cassini}. This result first arose from the equivalence of the theory with a scalar field model \cite{analogy}, which led to criticism from several fronts \cite{PPN2}; however, a later study implied that the result $\ga = 1/2$ could be obtained directly from the original $f(R)$ theory \cite{chiba} (see Ref. \cite{Faraoni} for a follow-up and criticism).

Despite the significant literature on these $f(R)$ models, another interesting possibility has been neglected until recent times: including not only a non-minimal scalar curvature term in the Einstein-Hilbert Lagrangian density, but also a non-minimum coupling between the scalar curvature and the matter Lagrangian density; indeed, these are only implicitly related in the action functional, since one expects that covariantly invariant terms in ${\cal L}$ should be constructed by contraction with the metric ({\it e.g.} the kinetic term of a real scalar field, $g^\mn \chi_{,\mu} \chi_{,\nu}$). In regions where the curvature is high (which, in GR, are related to regions of high energy density or pressure), the implications of such theory could deviate considerably from those predicted by Einstein's theory \cite{Lobo}. Related proposals have been put forward previously to address the problem of the accelerated expansion of the Universe \cite{expansion} and the existence of a cosmological constant \cite{cosmological}. Other studies have studied the behaviour of matter, namely changes to geodetic behaviour \cite{Lobo}, the possibility of modelling dark matter \cite{dark matter}, the violation of the highly constrained equivalence principle \cite{equivalence}\footnote{The connection between the equivalence principle and the interaction between dark energy and dark matter has been discussed in Ref. \cite{equivalence2}.} and the effect on the hydrostatic equilibrium of spherical bodies such as the Sun; also, a viability criterion for these generalized $f(R)$ theories has been obtained \cite{viability}.

In this study, we focus on the equivalence of a theory displaying a non-minimal coupling of the scalar curvature with matter with a scalar-tensor theory; through a conformal transformation of the metric, this yields a purely scalar theory, that is, one in which the curvature term appears isolated from any scalar field contribution. For definitiveness, we recast the theory in a form that is as consistent as possible with the work of Ref. \cite{damour}, and in close analogy with the available equivalence with $f(R)$ models \cite{analogy}.

This work is divided into the following sections: first, we introduce the gravity model and discuss some of its features; then, we recast it as a scalar-tensor theory with a suitable dynamical identification of the scalar fields, and then as a scalar theory with a conformally related metric and redefined scalar fields. The later prompts for a computation of the PPN parameters $\be$ and $\ga$, which is followed by a discussion of our results.

\section{The model}

Following the discussion of the previous section, one considers the following action \cite{Lobo},

\beq S = \int \left[ \ka f_1(R) + f_2(R) {\cal L} \right] \sqrt{-g}~ d^4 x ~~, \label{action} \eeq

\noindent where $\ka = c^4 /16 \pi G$, $f_i(R)$ (with $i=1,2$) are arbitrary functions of the scalar curvature $R$, ${\cal L}$ is the Lagrangian density of matter and $g$ is the metric determinant; the metric signature is $(-,+,+,+)$. The standard Einstein-Hilbert action is recovered by taking $f_2=1$ and $f_1= R - 2 \La$,  and $\La$ is the cosmological constant (from now on, one works in a unit system where $c= \ka = 1$).

Variation with respect to the metric $g_\mn$ yields the modified Einstein equations of motion, here arranged as

\beqa && \label{EE} \left( F_1 + F_2 {\cal L} \right) \left(R_\mn - {1 \over 2} g_\mn R \right) = 8\pi G f_2 T_\mn + \\ \nonumber && {1 \over 2} \left[f_1 - \left(F_1 + F_2  {\cal L} \right)R \right] g_\mn +  \left( \square_\mn - g_\mn \square \right) \left(F_1 + F_2 {\cal L} \right) ~~, \eeqa

\noindent where one defines $\square_\mn \equiv \nabla_\mu \nabla_\nu$ for convenience, as well as $F_i(R) \equiv df_i(R) / dR$, omitting the argument. The matter energy-momentum tensor is, as usually, defined by

\beq T_\mn = -{2 \over \sqrt{-g}} {\de \left(\sqrt{-g} {\cal L} \right) \over \de g^\mn } ~~. \eeq

The Bianchi identities, together with the identity $\left( \square \nabla_\nu - \nabla_\nu \square \right)F_i = R_\mn \nabla^\mu F_i$, imply the non-(covariant) conservation law

\beq \label{cons} \nabla^\mu T_\mn = { F_2 \over f_2} \left( g_\mn {\cal L} - T_\mn \right) \nabla^\mu R ~~, \eeq

\noindent and, as expected, in the  limit $f_2(R) = 1$, one recovers the conservation law $\nabla^\mu T_\mn = 0$. 

Since the energy-momentum tensor is not covariantly conserved, one concludes that the motion of matter distribution characterized by a Lagrangian density ${\cal L}$ is non-geodesical. This, of course, may yield a violation of the Equivalence principle, if the right-hand side of Eq. (\ref{cons}) differs between distinguishable matter distributions, which could be used to experimentally test the model and obtain constraints on $\lambda$.

\subsection{Equivalence with scalar-tensor theory}

\subsubsection{Survey}

As in usual $f_1(R)$ models, one may rewrite the considered mixed curvature model as a scalar-tensor theory. One wishes to establish an equivalent action in the Jordan frame (where the scalar curvature appears coupled linearly to a function of scalar fields); this may be obtained in a similar fashion to usual $f(R)$ models (with $f_2(R) = 1$), so that the equations of motion derived from the action functional coincide with those derived directly from the action Eq. (\ref{action}). However, one cannot simply consider an action of the form

\beq S = \int \left[ g(\Phi)R -V(\Phi) + h(\Phi) {\cal L} \right] \sqrt{-g}~ d^4 x  \label{actionone}~~. \eeq

\noindent The equivalence with the action (\ref{action}) must stem from the equations of motion obtained through the variation of with respect to the non-dynamical scalar field $\Phi$; through a suitable definition of the functions $g(\Phi)$, $h(\Phi)$ and the potential $V(\Phi)$, the variation of the action (\ref{actionone}) should yield the dynamic identification $\Phi = \Phi(R)$, so that both actions are equal, that is

\beqa g(\Phi(R))R -V(\Phi(R))  & = & f_1(R) ~~,\\ \nonumber h(\Phi(R)) & = & f_2(R) ~~.\eeqa

\noindent or, conversely, 

\beqa \label{converse} g(\Phi) R(\Phi) - V(\Phi) & = & f_1(R(\Phi)) ~~,\\ \nonumber h(\Phi) & = & f_2(R(\Phi)) ~~. \eeqa

\noindent From action (\ref{actionone}), the equation of motion for $\Phi$ reads

\beq \label{eqmotionone} g'(\Phi) R -V'(\Phi) + h'(\Phi){\cal L} =0 ~~, \eeq 

\noindent where the prime denotes differentiation with respect to $\Phi$. Integrating, one gets

\beq g(\Phi) R -V(\Phi) + h(\Phi){\cal L} = f(R) ~~, \eeq 

\noindent which then yields an action without matter content,

\beq S = \int f(R) \sqrt{-g}~ d^4 x ~~, \eeq

\noindent indicating that one can resort to just one scalar field only when dealing with pure $f(R)$ models (that is, with $f_2(R) = 1$). This pathology has already been pointed out in Ref. \cite{viability}, and in a following study \cite{Sotiriou} it is shown that one must resort to a second auxiliary real scalar field $\psi$. In this approach, the authors chose an action of the form

\beq S= \int \left[ f_1(\phi) + f_2(\phi) {\cal L} + \psi (R - \phi) \right] \sqrt{-g}~d^4 x ~~, \label{actiontwo} \eeq

\noindent so that the second scalar field acts as a Lagrange multiplier, enforcing the identification $\phi = R$; however, variation of the above with respect fo $\phi$ yields the relation

\beq \psi = F_1(\phi) + F_2(\phi) {\cal L}~~, \label{psi} \eeq

\noindent so that the two scalar fields are independent, if ${\cal L} \neq 0$ or $F_2 \neq 0$ (with $F_2=0$ falling back to the trivial case $f_2=1$ and $\psi = F_1(\phi)$). The on-shell action obtained by replacing Eq. (\ref{psi}) into action (\ref{actiontwo}) is given by

\beqa \label{actionone1} && S = \int \sqrt{-g}~d^4 x \times \\ \nonumber && \left[ f_1(\phi) + F_1(\phi) (R - \phi) + \left[ f_2(\phi) +F_2(\phi)(R-\phi)\right] {\cal L} \right] \ ~~. \eeqa

\noindent This differs from the unobtainable form of Eq. (\ref{actionone}) in the dependence of $F(\phi, {\cal L}) = F_1(\phi) + F_2(\phi) {\cal L}$ on both the scalar field $\phi$ and the Lagrangian matter density.

As in usual $f(R)$ models, a suitable conformal transformation to the metric $g^*_\mn = F(\phi,{\cal L}) g_\mn$ could be used to transform the above action of Eq. (\ref{actionone1}) into a functional where the scalar curvature appears decoupled from other fields. However, the presence of the Lagrangian density in this transformation would render any comparison with standard scalar theories too complicated (including the extraction of the PPN parameters $\be$ and $\ga$); for this reason, the second scalar field $\psi$ will be kept for the remainder of this study.

Notice also that one also could opt for the equivalence with a theory with just one scalar field, by taking the action

\beq \label{actionone2} S= \int \left[ f_1(\phi) + F_1(\phi) (R - \phi) +  f_2(R) {\cal L} \right] \sqrt{-g}~d^4 x ~~,\eeq

\noindent However, the latter is not written in the Jordan frame, as the scalar curvature appears in a non-linear fashion.

\subsubsection{Two-field scalar-tensor model}

In this study, one establishes the equivalence between action (\ref{action}) and a scalar-tensor theory through Eq. (\ref{actiontwo}), here rewritten as a Jordan-Brans-Dicke theory with a potential,

\beq S= \int \left[ \psi R - V(\phi,\psi) + f_2(\phi) {\cal L} \right] \sqrt{-g}~d^4 x ~~, \label{STaction} \eeq

\noindent where $\phi$ and $\psi$ are scalar fields and one defines 

\beq V(\phi,\psi) = \phi \psi- f_1(\phi) ~~. \eeq

As discussed in the previous section, variation of the metric with respect to both scalar fields yields the dynamical equivalence $ \phi = R $ and $\psi = F_1(\phi) + F_2(\phi){\cal L} $. Substituting into Eq. (\ref{STaction}), one recovers the action for the mixed curvature model, Eq. (\ref{action}).

With the above taken into consideration, one should notice that the second scalar field $\psi$ is not a function of the curvature (or $\phi$) alone, but also of the matter Lagrangian ${\cal L}$ (which is itself a scalar). The degrees of freedom of the matter fields $\chi$ which appear in the Lagrangian ${\cal L}(g^\mn,\chi)$ (including any kinetic terms) are displayed in the Einstein field equations, through the corresponding energy-momentum tensor $T_\mn$:

\beqa & & R_\mn - {1 \over 2} g_\mn R = 8 \pi G {f_2(\phi) \over \psi} T_\mn +  \\ \nonumber &&  -{1 \over 2} g_\mn {V(\phi,\psi) \over \psi}  + {1 \over \psi} \left( \square_\mn - g_\mn \square \right) \psi  ~~, \label{STEE} \eeqa

\noindent which, introducing $\phi = R $ and $\psi = F_1(\phi) + F_2(\phi){\cal L} $, recovers Eq. (\ref{EE}).

Using the Bianchi identities and the previous relation, one obtains

\beqa && \nabla^\mu T_\mn = \\ \nonumber && {1 \over f_2(\phi)} \left[ \nabla_\nu V(\phi,\psi) - R \nabla_\nu \psi - F_2(\phi) T_\mn \nabla^\mu \phi \right] ~~. \eeqa

\noindent Since 

\beq \nabla_\nu V(\phi,\psi) = \left[ \psi - F_1(\phi) \right] \nabla_\nu \phi + \phi \nabla_\nu \psi \eeq

\noindent  we get

\begin{widetext} \beq \nabla^\mu T_\mn = {1 \over f_2(\phi)} \left[ \left(\phi- R \right) \nabla_\nu \psi +\left[ \left( \psi - F_1(\phi) \right)g_\mn - F_2(\phi) T_\mn \right] \nabla^\mu \phi \right]  ~~, \eeq \end{widetext}

\noindent which, upon the substitution, $\psi = F_1(\phi) + F_2(\phi){\cal L}$ and $\phi= R$, collapses back to Eq. (\ref{cons}).

\section{Equivalence with a scalar theory}

One may now perform a conformal transformation (see {\it e.g.} Ref. \cite{conformal}), so that the curvature appears decoupled from the scalar fields $\phi$, $\psi$ (yielding the action in the so-called Einstein frame), by writing $g^*_\mn = \psi g_\mn =A^{-2}(\psi) g_\mn$, with $A(\psi) = \psi^{-1/2}$, one obtains

\beqa \sqrt{-g} & = & \psi^{-2}\sqrt{-g^*}~~, \\ \nonumber  R & = & \psi\left[R^* - 6\sqrt{\psi} \square^*\left({1\over \sqrt{\psi}} \right) \right]~~,\eeqa

\noindent where $\square^*$ denotes the D'Alembertian operator, defined from the metric $g^*_\mn$. From the definition of the energy-momentum tensor, this implies that $T^*_\mn = A^2(\psi) T_\mn$.

Introducing the above into the action (\ref{STaction}) yields

\beqa \label{action*} && S = \int  \sqrt{-g^*} ~d^4 x \times \\ \nonumber && \left[ R^* - 6\sqrt{\psi} \square^* \left({1\over \sqrt{\psi}} \right)  - 4 U + f_2 A^4 {\cal L} (A^2 g^*_\mn,\chi) \right] ~~,\eeqa

\noindent where one defines 

\beq U(\phi,\psi) = A^4(\psi) V(\phi,\psi)/4 = {\phi \over 4\psi} - {f_1(\phi) \over 4\psi^2}~~,\eeq

\noindent Notice that there are two couplings between the scalar fields $\phi$, $\psi$ and matter: the explicit coupling given by the factor $f_2(\phi)A^2(\psi)$, and an additional coupling due to the rewriting of the metric (in the Jordan frame) $g_\mn$ in terms of the new metric $g^*_\mn$.

One now attempts to recast the action in terms of two other scalar fields, endowed with a canonical kinetic term. For this, one first integrates the covariant derivative term by parts and uses the metric compatibility relations, obtaining

\beqa && -6 \int \sqrt{\psi} \square^* \left( {1\over \sqrt{\psi}} \right) \sqrt{-g^*}~d^4x= \\ \nonumber && -6 \int \nabla^*_\mu \left[ \sqrt{\psi} \nabla^{*\mu} \left( {1\over \sqrt{\psi}} \right) \sqrt{-g^*} \right]~d^4x + \\ \nonumber &&  6\int \nabla^*_\mu \sqrt{\psi} \nabla^{*\mu} \left( {1\over \sqrt{\psi}} \right)  \sqrt{-g^*}~d^4x~~. \eeqa

\noindent By resorting to the divergence theorem, the first integral may be dropped, yielding

\beqa && -6 \int \sqrt{\psi} \square^* \left( {1\over \sqrt{\psi}} \right) \sqrt{-g^*}~d^4x = \\ \nonumber && -{3 \over 2}\int g^{*\mn} {F_{,\mu} F_{,\nu}\over F^2} \sqrt{-g^*}~d^4x ~~. \eeqa

One obtains the action 

\beqa && S = \int \sqrt{-g^*} ~d^4 x \times \\ \nonumber && \left[ R^* -{3 \over 2} g^{*\mn} {F_{,\mu} F_{,\nu}\over F^2}  - 4 U + f_2 A^4 {\cal L} (A^2 g^*_\mn,\chi) \right]  ~~,\eeqa

One may redefine the two scalar fields, so that their kinetic terms may be recast in the canonical way. Specifically, one aims at writing (see Ref. \cite{damour}):

\beq {3 \over 2} {\psi_{,\mu} \psi_{,\nu}\over F^2} = {3 \over 2} \left(\log~\psi\right)_{,\mu} \left(log~\psi \right)_{,\nu}  \equiv 2 \sigma_{ij} \varphi^i_{,\mu} \varphi^j_{,\nu} ~~,\eeq

\noindent with $i,~j=1,~2$; $\sigma_{ij}$ is the metric of the two-dimensional space of scalar fields (field-space metric, for short), and $\varphi^1$, $\varphi^2$ the two new scalar fields. Clearly, this kinetic term prompts for the identification

\beq \label{deffields} \varphi^1 = {\sqrt{3} \over 2}\log~\psi ~~~~,~~~~\varphi^2 = \phi ~~, \eeq

\noindent and the field metric

\beq \si_{ij} = \left(\begin{array}{cc}1 & 0 \\0 & 0\end{array}\right) ~~,\eeq

\noindent which indicates that only $\varphi^1$ has a kinetic term. Despite this, it is clear that $\varphi^2 = \phi$ is a distinct degree of freedom, since one cannot rewrite the potential $U(\phi,\psi)$ in terms of one scalar field alone:

\beqa && U(\varphi^1, \varphi^2) ={1\over 4} \exp\left(-{2 \sqrt{3}\over 3}\varphi^1\right) \times\\ \nonumber &&  \left[ \varphi^2 - f_1(\varphi^2) \exp\left(-{2 \sqrt{3}\over 3}\varphi^1\right) \right]~~. \eeqa

\noindent In the trivial case $f_2(R) = 1$ or ${\cal L} = 0$, one gets $\psi = F_1(\phi) $, that is, $ \varphi^1 \propto \log F_1(\varphi^2)$, so that one degree of freedom is lost and this potential may be written as a function of just one the fields.

Since the inverse field metric $\si^{ij}$ will be required to raise latin indexes throughout the text, one still has do deal with the particular form of $\si_{ij}$, which is non-invertible. In order to cope with this caveat, one is free to add an antisimmetric part, rewriting it as

\beq \si_{ij} = \left(\begin{array}{cc}1 & a \\-a & 0\end{array}\right) ~~,\eeq

\noindent with inverse

\beq  \si^{ij} = a^{-1} \left(\begin{array}{cc}0 & -1 \\1 & a^{-1} \end{array}\right) ~~.\eeq

\noindent Clearly, no physical results can depend on the value $a$ that shows in the off-diagonal part of $\si_{ij}$ -- in particular, the values of the PPN parameters $\be$ and $\ga$, as shall be shown.

With this choice, the action now reads

\beqa \label{actionfinal} S & = & \int  \sqrt{-g^*} ~d^4 x \times \\ \nonumber && \left[  R^* -2 g^{*\mn} \si_{ij} \varphi^i_{,\mu}\varphi^j_{,\nu} -4 U(\varphi^1,\varphi^2) + f_2(\varphi^2){\cal L}^* \right] ~~. \eeqa

\noindent where one defines ${\cal L}^* =  A^4 (\varphi^1) {\cal L}$, with $A(\varphi^1) = \exp \left(-(\sqrt{3}/3)\varphi^1 \right)$. By varying the action with respect to the metric $g^{*\mn}$, one obtains

\beqa && R^*_\mn - {1 \over 2} g^*_\mn R^* = 8 \pi G f_2(\varphi^2) T^*_\mn + \\ \nonumber && \si_{ij} \left( 2\varphi^i_{,\mu}\varphi^j_{,\nu} - g^*_\mn g^{*\al\be} \varphi^i_{,\al}\varphi^j_{,\be} \right)- 2 g^*_\mn U  ~~. \eeqa

Defining $T^* = g^{*\mn} T^*_\mn$ and $\al_i = A^{-1} (\partial A/\partial \varphi^i) $, so that

\beq \al_1 = -{\sqrt{3} \over 3}~~~~,~~~~ \al_2 = 0~~. \eeq

\noindent one obtains, after some algebra

\beq {\de \left( \sqrt{-g^*} f_2 {\cal L}^* \right)\over \sqrt{-g^*}}= - {\sqrt{3} \over 3} f_2 T^* \de \varphi^1 + F_2 {\cal L}^* \de \varphi^2  ~~, \label{var} \eeq

\noindent so that the Euler-Lagrange equation for each field $\varphi^i$ read

\beqa a \square^* \varphi^1 & = & - B_2 + 4 \pi G F_2 {\cal L}^*  ~~,  \\ \nonumber a \square^* \varphi^2 &=& B_1 + {B_2 \over a} + {4 \pi G  \sqrt{3} \over 3a}  f_2 T^* - {4 \pi G  \over a} F_2 {\cal L}^*  ~~, \label{EL1} \eeqa

\noindent where one defines $B_i = \partial U/ \partial \varphi^i$.

Alternatively, one may use the field-space metric $\si_{ij}$ to raise and lower latin indexes, so that $\al^i = \si^{ij} \al_j$ and $B^i = \si^{ij}B_j$, rewriting Eq. (\ref{var}) as

\beq {\de \left( \sqrt{-g^*} f_2{\cal L}^* \right) \over \sqrt{-g^*}} = \left( \al_i f_2 T^* + \de_{i2} F_2  {\cal L}^* \right) \de \varphi^i ~~, \eeq

\noindent and Eq. (\ref{EL1}) becomes

\beq \square^* \varphi^i = B^i - 4 \pi G \al^i f_2 T^* - 4 \pi G \si^{i2} F_2 {\cal L}^* ~~. \eeq

This form enables a prompt comparison with the $f_2 = 1$ limit, yielding the usual expression \cite{PPN} for just one scalar field, $ \square^* \varphi = B - 4 \pi G \al T^* $ (with $ \varphi = \varphi^1$ and $\al = \al_1$).

Using the Bianchi identities, the expression for the non-covariant conservation of the energy-momentum tensor in the Einstein frame is also attained:

\beq \label{noncons} \nabla^{*\mu} T^*_\mn =  -{\sqrt{3} \over 3} T^* \varphi^1_{,\nu} + {F_2 \over f_2} \left( g^*_\mn {\cal L}^* - T^*_{\mn}  \right) \nabla^\mu \varphi^2 ~~.  \eeq

\noindent If, for consistency, one rewrites this in the Jordan frame, the conformal transformation properties of the contravariant derivative eliminate the first term in the {\it r.h.s.}, and the substitution $\varphi^2 = \phi = R$ yields Eq. (\ref{cons}). Again, taking $f_2(R) = 1$ one recovers $\nabla^{*\mu} T^*_\mn = \al T^* \varphi_{,\nu}$ (or, in the Jordan frame, the covariant conservation law $\nabla^\mu T_\mn = 0$).

\section{Parameterized Post-Newtonian formalism}

Assuming that the effect of the non-minimum coupling of curvature to matter is perturbative, one may write $ f_2(R) = 1 + \la \de_2(R) $, with $\la \de_2 \ll 1$, so that the current bounds on the equivalence principle are respected. Substituting into Eq. (\ref{noncons}) one gets, at zeroth-order in $\la$,

\beq \nabla^{*\mu} T^*_\mn \simeq \al_j T^* \varphi^j_{,\nu} ~~, \eeq

\noindent which amounts to ignoring the $f_2(\varphi^2)$ factor in the action (\ref{actionfinal}), so that $f_2$ manifests itself only through the coupling $A^2(\varphi^1)$ present in $T^*$, and the derivative of $\varphi^1$ (since $\varphi^1 \propto \log \psi$ and $\psi = F_1 + F_2 {\cal L}$). In this case, the matter action reads

\beq S_m = \int A^4(\psi) {\cal L} (A^2(\psi) g^*_\mn, \chi) \sqrt{-g^*} d^4x~~.\eeq

If there is no characteristic length ruling the added gravitational interaction (that is, both scalar fields are light, leading to long range interactions), this manifestation of a ``physical metric'' in the matter action allows one to resort to calculate the PPN parameters $\be$ and $\ga$ \cite{damour}: 

\beq \be - 1 = {1 \over 2} \left[ {\al^i \al^j \al_{j,i} \over \left( 1 + \al^2 \right)^2 } \right]_0~~~~,~~~~ \ga - 1 = -2 \left[ {\al^2 \over 1+ \al^2 } \right]_0 ~~, \label{PPNparameters} \eeq

\noindent where $\al_{j,i} = \partial \al_j / \partial \varphi^i$ and $\al^2  = \al_i \al^i = \si^{ij} \al_i \al_j $; the subscript $_0$ indicates that the quantities should be evaluated at their asymptotic values $\varphi^i_0$, related with the cosmological values of the curvature and matter Lagrangian density (as shall be further discussed in the following section). 

Since the $\al_i$'s are constant for $ i= 1,~2$, one gets that $\al_{j,i}= 0$; hence, the PPN parameter $\beta$ is unitary. Moreover, since $\al_i = (\al_1,0)$ is orthogonal to $\al^i = (0, \al_1 / a )$, one gets $\al^2 = 0$. Hence, the PPN parameter $\ga$ is also unitary. As argued previously, these results do not reflect the particular value of $a$ chosen in the antisymmetric part of $\si_{ij}$.

In light of the overall discussion presented, it is clear that the two degrees of freedom embodied in the two independent scalar fields $\varphi^1$ and $\varphi^2$ stem not only from the non-minimal curvature term in action (\ref{action}), but also of the non-minimal coupling of $R$ to the matter Lagrangian density. Given this, it is clear that a ``natural'' choice for the two scalar fields (in the Jordan frame) would be $\phi= R$ and $\psi = {\cal L}$: this more physical interpretation {\it ab initio} comes at the cost of more evolved calculations, since both redefined scalar fields in the Einstein frame will depend on $\phi$ and $\psi$. This less pedagogical approach is deferred to Appendix \ref{app1}.

\section{Discussion and Conclusions}

The result $\ga = 1$ is key to our study: it is clear that, in the standard $f_2(R) = 1$ theories, $\al^2 $ does not vanish (it is a purely algebraic, not matricial result and $\al^2 = 1/3$), and the resulting PPN parameter $\ga = 1/2$, which violates well-known observational bounds! A more thorough discussion on the ongoing debate concerning the value of the PPN parameter $\ga$ for $f(R)$ theories is deferred to Appendix \ref{app2} -- with special focus to what is believed to be a misconception in the identification of the equivalence with a scalar-tensor theory.

In the $f_2(R) \neq 1$ case, the added degree of freedom that a non-minimal coupling of curvature to matter implies yields not one, but two scalar fields: as a result, a two-dimensional field-space metric $\si_{ij}$ arises; from the redefinition of the fields necessary to absorb non-canonical kinetic terms after the conformal transformation to the Einstein frame, it follows that this enables a vanishing $\al^2 = \al_i \al_j \si^{ij}$ term, yielding no post-Newtonian observational signature that discriminates these models from General Relativity. However, this conclusion is valid only in zeroth-order in $\la$: if more terms are allowed, the non-covariant conservation law for the energy-momentum tensor is no longer of the form treated in Ref. \cite{damour}, and more elaborate calculations would have to be performed in order to extract the PPN parameters $\be$ and $\ga$.

It is important to highlight that a naive analysis of the model under scrutiny might predict no difference between the PPN parameters arising from the trivial $f_2(R) = 1$ and the non-minimal $f_2(R) \neq 1$ cases: indeed, since this function is coupled to the matter Lagrangian density one might expect that, outside of the matter distribution (${\cal L}=0$), the theory would collapse back to the usual $f(R)$ scenario. However, this misinterpretation is resolved by Eq. (\ref{PPNparameters}), which is evaluated at the cosmological values $\varphi^i_0$: for this reason, the relevant value for ${\cal L}$ is given by the Lagrangian density of the overall cosmological fluid, not of the local environment. The issue of suitably identificating this contribution is discussed in Ref. \cite{BLP}.

Finally, note that the result $\be = \ga \simeq 1$ is approximate, since it corresponds to dropping the term $f_2(\varphi^2)$ in action (\ref{actionfinal}): a future work should consider a perturbative approach, which could perhaps yield a small, $\la$-dependent deviation from unity, thus marking a clear (even if small) departure between the model studied here and General Relativity.


\appendix

\section{Alternative formulation of the two-field equivalent scalar-tensor theory}
\label{app1}

We present here an alternative formulation where one chooses instead to express action (\ref{action}) through the equivalent expression

\beq \label{STactionalt}  S=  \int \left[ F(\phi,\psi) R - V (\phi,\psi)+ f_2(\phi) {\cal L}(g_\mn,\chi) \right] \sqrt{-g}~ d^4 x ~~,\eeq

\noindent where one defines 

\beqa F(\phi,\psi) & = & F_1(\phi) + F_2(\phi)\psi~~, \nonumber \\ V(\phi,\psi) & = &  \phi F(\phi,\psi) - f_1(\phi)~~. \eeqa

One obtains the equivalence with the model under scrutiny by writing the equations of motion for the scalar fields,

\beqa \label{eqmotionscalaralt} F_2(\phi) \left( {\cal L} -  \psi \right) + \left[ F_1'(\phi) + F_2'(\phi) \psi \right] (R - \phi) = 0 ~~, \\ \nonumber   F_2(\phi) (R - \phi) = 0 ~~,  \eeqa

\noindent implying that $ \phi = R $ (or $F_2(\phi) = 0 \rightarrow f_2 = 1$, the trivial result) and, therefore, $\psi = {\cal L} $ . Substituting into Eq. (\ref{STactionalt}), one recovers the action for the mixed curvature model, Eq. (\ref{action}).

It should be stressed that the GR limit $f_1 = R$ and $f_2 = 1$ disables the identification $\phi = R$: one may write $f_1 (\phi) = \phi + \ep \de_1(\phi)$ and $f_2 (\phi) = 1 + \la \de_2(\phi)$, so that Eqs. (\ref{eqmotionscalaralt}) become

\beqa \label{eqmotionscalar2} \la \de_2'(\phi) \left( {\cal L} -  \psi \right) + \left[ \ep \de_1''(\phi) + \la \de_2''(\phi) \psi \right] (R - \phi) = 0 ~~, \\ \nonumber   \la \de'(\phi) (R - \phi) = 0 ~~,  \eeqa

\noindent and taking the limit $\ep_i \rightarrow 0$ gives a trivial identity. For this reason, it is misleading to insert the above approximations in results that stem from the scalar field approach to $f(R)$ theories, and only then consider the GR limit: the formalism itself breaks down at its inception. For this reason, one concludes that one cannot simply take the limit $\ep \rightarrow 0$ and argue that, as a $f(R)$ theory collapses back to GR, so should the PPN parameter $\ga$ approach unity, which does not happen if $\ga = 1/2$ and does not show a dependence on $\ep$.

Variation of the action (\ref{STactionalt}) with respect to the metric yield the Einstein equations,

\beqa & & F(\phi,\psi) \left( R_\mn - {1 \over 2} g_\mn R \right)  = 8 \pi G f_2(\phi) T_\mn   \\ \nonumber &&  -{1 \over 2} g_\mn V(\phi,\psi)  + \left( \square_\mn - g_\mn \square \right) F(\psi,\phi)  ~~, \label{STEEalt} \eeqa

\noindent which, introducing $\phi = R $ and $\psi = {\cal L} $, recovers Eq. (\ref{EE}).

Using the Bianchi identities and the previous relation, one obtains

\beqa && \nabla^\mu T_\mn = \\ \nonumber && {1 \over f_2(\phi)} \left[ \nabla_\nu V(\phi,\psi) - R \nabla_\nu F(\phi,\psi) - F_2(\phi) T_\mn \nabla^\mu \phi \right] ~~. \eeqa

\noindent Since 

\beqa && \nabla_\nu V(\phi,\psi) = \\ \nonumber && \left[ \psi F_2(\phi) + \phi \left( F_1'(\phi) + \psi F_2'(\phi) \right) \right] \nabla_\nu \phi + \phi F_2(\phi) \nabla_\nu \psi \eeqa

\noindent and

\beqa && \nabla_\nu F(\phi,\psi) = \\ \nonumber &&  \left( F_1'(\phi) + F_2'(\phi) \psi \right) \nabla_\nu \phi + F_2(\phi) \nabla_\nu \psi~~, \eeqa

\noindent  we get

\begin{widetext} \beq \nabla^\mu T_\mn = {1 \over f_2(\phi)} \left[ F_2(\phi) \left( g_\mn \psi - T_\mn \right) \nabla^\mu \phi + \left( \left[F_1'(\phi) + \psi F_2'(\phi) \right] \nabla_\nu \phi + F_2(\phi) \nabla_\nu \psi \right) (\phi - R)  \right] ~~, \eeq \end{widetext}

\noindent which, upon the substitution $\psi =  {\cal L}$ and $\phi= R$, collapses back into Eq. (\ref{cons}).

\subsection{Equivalence with a scalar theory}

In this alternative formulation, the adequate conformal transformation to decouple the scalar curvature from the scalar fields $\phi$, $\psi$ is given by $g^*_\mn = F(\phi,\psi) g_\mn =A^{-2}(\phi,\psi) g_\mn$, with $A(\phi,\psi) = F^{-1/2}(\phi,\psi)$. One obtains

\beqa \sqrt{-g} & = & F^{-2}\sqrt{-g^*}~~, \\ \nonumber  R & = & F\left[R^* - 6\sqrt{F} \square^*\left({1\over \sqrt{F}} \right) \right]~~.\eeqa

Introducing the above into the action (\ref{STactionalt}) yields

\beqa \label{action*alt} && S = \int  \sqrt{-g^*} ~d^4 x \times \\ \nonumber && \left[ R^* - 6\sqrt{F} \square^* \left({1\over \sqrt{F}} \right)  - 4 U + f_2 A^4 {\cal L} (A^2 g^*_\mn,\chi) \right] ~~,\eeqa

\noindent where one defines $U(\phi,\psi) = A^4(\phi,\psi) V(\phi,\psi)/4$.

One now attempts to recast the action in terms of two other scalar fields, endowed with canonical kinetic term. For this, one repeats the integration of the covariant derivative term by parts and uses the metric compatibility relations, finally obtaining the action 

\beqa && S = \int \sqrt{-g^*} ~d^4 x \times \\ \nonumber && \left[ R^* -{3 \over 2} g^{*\mn} {F_{,\mu} F_{,\nu}\over F^2}  - 4 U + f_2 A^4 {\cal L} (A^2 g^*_\mn,\chi) \right]  ~~,\eeqa

\noindent As one aims to allow for an immediate comparison with the $f_2 = 1$ scenario, it is interesting to isolate the contributions arising from the non-minimal scalar curvature term and from its coupling with matter as clearly as possible, namely

\beq log~F = log~\left(F_1 + F_2 \psi \right) = log~F_1 + log\left(1 + {\psi F_2 \over F_1} \right)~~,\eeq

\noindent assuming that $F_1 \neq 0$. If one defines 

\beq \label{deffieldsalt} \varphi^1 = {\sqrt{3} \over 2} log~F_1(\phi) ~~~~,~~~~\varphi^2 = {\sqrt{3} \over 2} log\left(1 + {\psi F_2(\phi) \over F_1(\phi)} \right) ~~, \eeq

\noindent then the comparison is transparent: when $f_2= 1$, the second scalar field vanishes, and the first scalar field coincides with the usual redefined scalar field $\varphi^1 = log ~F_1$ arising in $f_1(R)$ models (although in many studies some terms are overlooked, see Appendix \ref{app2}). Also, when $f_1 = R$, the first scalar field $\varphi^1$ vanishes.

For the particular choice of fields Eqs. (\ref{deffieldsalt}), one obtains cross-products between the derivatives of $\varphi^1$ and $\varphi^2$, so that the $\si_{ij}$ field-space metric displays non-vanishing off-diagonal elements; since one has absorved the numerical factors in the redifinition, this field-space metric is trivially given by $\si_{ij} = 1$ for $i,~j=1,2$. However, this field-space metric is not invertible, a problem already found in the main body of this work; as before, this issue may be surpassed by adding an antisymmetric part to this field metric, adopting instead the form

\beq \si_{ij} = \left(\begin{array}{cc}1 & 1 \\ 1 & 1\end{array}\right) + \left(\begin{array}{cc}0 & 1-a \\a-1 & 0\end{array}\right) =\left(\begin{array}{cc}1 & 2-a \\a & 1\end{array}\right) ~~,\eeq

\noindent with inverse

\beq  \si^{ij} = {1 \over (1-a)^2} \left(\begin{array}{cc}1 & a-2 \\-a & 1\end{array}\right) ~~.\eeq

\noindent As already discussed, the physical results will not depend on the value $a$.

With this choice, the action now reads

\beqa \label{actionfinalalt} S & = & \int  \sqrt{-g^*} ~d^4 x \times \\ \nonumber && \left[  R^* -2 g^{*\mn} \si_{ij} \varphi^i_{,\mu}\varphi^j_{,\nu} -4 U(\varphi^1,\varphi^2) +  f_2(\phi(\varphi^1)){\cal L}^* \right] ~~. \eeqa

\noindent where one defines ${\cal L}^* = A^4 (\varphi^1,\varphi^2) {\cal L}$ as before, but using the new redefined fields $\varphi^i$; notice that the non-minimal coupling is written in terms of $\varphi^1$ alone. The Einstein field equations are

\beqa && R^*_\mn - {1 \over 2} g^*_\mn R^* = 8 \pi G f_2 T^*_\mn + \\ \nonumber && 2\si_{ij} \varphi^i_{,\mu}\varphi^j_{,\nu} - g^*_\mn g^{*\al\be}\si_{ij} \varphi^i_{,\al}\varphi^j_{,\be} - 2 g^*_\mn U  ~~, \eeqa

\noindent and variation of the matter action with respect to each scalar field $\varphi^i$ yields, after some algebra

\beqa && { \de \left( f_2  \sqrt{-g^*} {\cal L}^* \right) \over \sqrt{-g^*}}= \\ \nonumber && \left[ f_2 \al_i T^* + F_2 \de_{i1}{\partial \phi \over \partial \varphi^1}   {\cal L}^* \right] \de \varphi^i  = \\ \nonumber && \left[ f_2 T^* - {2F_1 F_2 \over F_1'} \de_{i1}   {\cal L}^*\right] \al_i \de \varphi^i ~~,\eeqa

\noindent where, as before, one defines

\beq \al_i = {\partial~ log~A \over \partial \varphi^i} = -{1 \over 2}{\partial~ log~F \over \partial \varphi^i} = -{\sqrt{3}\over 3} \equiv \al~~.\eeq

\noindent and, given definitions Eqs. (\ref{deffieldsalt}),

\beq \label{partial} {\partial \phi \over \partial \varphi^1 }= {2 \sqrt{3 }\over 3} {F_1 \over F_1'} = - 2\al  {F_1 \over F_1'} ~~. \eeq

The Euler-Lagrange equation for each field $\varphi^i$ reads

\beq \square^* \varphi^i = B^i - 4 \pi G   \left[ f_2 T^* - {2F_1 F_2 \over F_1'} \de_{i1}   {\cal L}^*\right] \al^i ~~,\eeq

\noindent In the $f_2 = 1$ limit, one recovers $ \square^* \varphi = B - 4 \pi G \al T^* $, as before.

Using the Bianchi identities, the expression for the non-covariant conservation of the energy-momentum tensor is

\beq \label{nonconsalt} \nabla^{*\mu} T^*_\mn = \left[g_\mn T^* -{2F_1  \over  F_1'} {F_2 \over f_2} \de_{i1} \left( g^*_\mn  {\cal L}^* - T^*_\mn \right) \right]\al_i \nabla^\mu \varphi^i~~.  \eeq

\noindent As discussed regarding Eq. (\ref{noncons}), the first term of the {\it r.h.s} is elliminated in the inverse conformal transformation back to the Jordan frame, and one obtains an expression that collapses to Eq. (\ref{cons}), after a suitable manipulation.

\subsection{Parameterized Post-Newtonian formalism}

Once again, assuming a perturbative effect of $f_2(R) = 1 + \la \de_2(R)$, with $\la \de_2 \ll 1$ yields, to zeroth-order in $\la$

\beq \nabla^{*\mu} T^*_\mn \simeq \al_j T^* \varphi^j_{,\nu} ~~, \eeq

\noindent which amounts to ignoring the $f_2(\phi)$ factor in the action (\ref{actionfinalalt}), so that $f_2$ manifests itself only through the coupling $A^2(\phi,\psi)$; in this case, one may write

\beq {\cal L}^* \simeq A^4(\varphi^1,\varphi^2) {\cal L}~~, \eeq

The computation of the PPN parameters $\be$ and $\ga$ is very similar to the one already performed: since $\al_i$'s are constant for $ i= 1,~2$, one gets that $\al_{j,i}=0$ and $\beta = 1$. One also obtains

\beqa \al^2 &=& \sigma^{ab} \al_i \al_j = \\ \nonumber && { \al^2 \over (1-a)^2 } \left[\begin{array}{cc}1 & 1\end{array}\right] \left[\begin{array}{cc}1 & a-2 \\-a & 1\end{array}\right]\left[\begin{array}{c}1 \\ 1 \end{array}\right] = 0~~, \eeqa

\noindent independently of the value $a$, as already discussed. Hence,  one confirms that $\ga =1$. As required, the particular choice of field-metric $\si$ and scalar fields (initially in the Jordan frame) $\phi$ and $\psi$ do not change the obtained results $\be = \ga = 1$.


\section{Discussion of previous results}
\label{app2}

In this appendix one addresses the issue of the $f_2(R) = 1$, $f_1(R) = f(R)$ case, very much discussed in the literature (see {\it e.g.} Ref. \cite{f(R)}) -- in particular, the claim that $f(R)$ models characterized by a light scalar field still allow for the PPN parameters $\be$ and $\ga$ to be close to unity. The action for this model is, in the Jordan frame, given by

\beq S = \int \left[ f(R) + {\cal L}(g_\mn,\chi) \right] \sqrt{-g}~ d^4 x ~~, \eeq

\noindent which is dynamically equivalent to 

\beq S = \int \left[ F(\phi) R - V (\phi) + {\cal L}(g_\mn,\chi) \right] \sqrt{-g}~ d^4 x ~~,\eeq

\noindent with the usual identification $\phi = R$, and the definition $ V(\phi) = \phi F(\phi) - f(\phi) $.

A conformal transformation $g^*_\mn = F g_\mn = A^{-2}g_\mn$, with $A(\phi) = F^{-1/2}(\phi)$, yields the action, in the Einstein frame,

\beqa && S =  \int \sqrt{-g^*} ~d^4 x \times \\ \nonumber &&  \left[ R^* -{3 \over 2} g^{*\mn} {F_{,\mu} F_{,\nu}\over F^2}   - 4 U  + A^4 {\cal L} (A^2 g^*_\mn,\chi) \right] ~~,\eeqa

\noindent where $U(\phi) = A^4(\phi) V(\phi)/4$.

One may redefine the scalar field as

\beq \varphi= {\sqrt{3} \over 2} ~log~F(\phi) = - \sqrt{3}~\log~A~~,\eeq

\noindent so that the action becomes

\beqa && S =  \int  \sqrt{-g^*} ~d^4 x \times \\ \nonumber && \left[ R^* -2 g^{*\mn} \varphi_{,\mu} \varphi_{,\nu}   - 4 U + A^4{\cal L} (A^2 g^*_\mn,\chi) \right] ~~,\eeqa

\noindent and the matter action depends not on the Einstein metric $g^*_\mn$, but on the original Jordan metric $g_\mn = A^2 g^*_\mn$:

\beqa S_m &= & \int A^4 {\cal L} (A^2 g^*_\mn, \chi) \sqrt{-g^*} d^4x = \\ \nonumber && \int {\cal L} (g_\mn, \chi) \sqrt{-g} d^4x ~~.\eeqa

One obtains

\beq \al = {\partial ~log~A \over \partial \varphi} = -{1 \over 2} {\partial ~log~F \over \partial \varphi} = -{\sqrt{3} \over 3} ~~, \eeq

\noindent identical to the previous result for $\al_a$. However, in this case the field-space metric $\si$ is one dimensional, and simply given by $\si_{11} = \si^{11} = 1$, so that $\al^2 = 1/3$. One obtains $\al_{,\varphi} = 0$, so that 

\beqa \be - 1& = & {1 \over 2} \left[ {\al^2 \al_{,\varphi} \over \left( 1 + \al^2 \right)^2 } \right]_0 = 0 \rightarrow \be = 1~~, \\ \nonumber \ga - 1 &= & -2 \left[ {\al^2 \over 1+ \al^2 } \right]_0 = -{1 \over 2} \rightarrow \ga = {1 \over 2} ~~,\eeqa

This indicates that general scalar-tensor theories with no {\it a priori} kinetic term for the long range scalar field (in the Jordan frame) are incompatible with observations. As the above example shows, the equivalence between $f(R)$ theories and such models falls within this category, and is therefore observationally ruled out. Furthermore, the result $\al^2 = 1/3$ enables, for the case of a single scalar field (see Ref. \cite{damour}), the identification of the Brans-Dicke coupling parameter 

\beq 2 \om +3 ={1 \over \al^2} \rightarrow \om = 0~~, \eeq 

\noindent which shows that $f(R)$ models may also be recast as a (sometimes used) generalized Jordan-Brans-Dicke model with no kinetic term,

\beq S =  \int \left[ \phi R -V(\phi) + {\cal L} \right] \sqrt{-g} d^4x~~,\eeq

\noindent with the dynamical identification $\phi = F(R) $ and a suitable potential $V(\phi)=R(\phi)F(R(\phi)) - f(R(\phi))$.

In several papers in the literature this equivalence with a scalar-tensor theory is given by the action

\beqa \label{actionsinglescalar} S & = & \int \sqrt{-g} ~d^4 x \times \\ \nonumber && \left[  F(\phi)R - Z(\phi) g^\mn \phi_{,\mu} \phi_{,\nu} - V(\phi) + {\cal L} (g_\mn,\chi) \right] ~~,\eeqa

\noindent with $Z(\phi) = 1$; for later convenience, one retains the kinetic function $Z(\phi)$. As discussed above, one can opt by an equivalent Jordan-Brans-Dicke theory with a scalar field dynamically identified through $\phi = F(R)$, and no kinetic term, that is, $\om = 0$.

It is easy to verify that variation of the action with respect to the scalar field $\phi$ will yield terms involving $Z'(\phi)$ (which vanishes, in the usual approach $Z(\phi)=1$ and the four-dimensional D'Alembertian operator, similarly to the classical Klein-Gordon equation. For this reason, the presence of a kinetic term in the above action implies that the dynamical identification $\phi = R$ (arising from the equation of motion of the scalar field $\phi$) fails.

Moreover, the above conformal transformation $g^*_\mn = F(\phi) g_\mn$ yields

\begin{widetext} 
\beq \label{actionZ} S =  \int \left[ R^* -{3 \over 2} g^{*\mn} {F_{,\mu} F_{,\nu}\over F^2} - g^{*\mn} Z(\phi) {\phi_{,\mu} \phi_{,\nu} \over F(\phi)} - 4 U(\phi) + A^4(\phi) {\cal L} (A^2(\phi) g^*_\mn,\chi) \right] \sqrt{-g^*} ~d^4 x ~~,\eeq
\end{widetext}

\noindent using the previous result relating $R^*$ and $R$, as well as $g^\mn = F(\phi) g^{*\mn}$ and $\sqrt{-g} =F^{-2}(\phi) \sqrt{-g^*} $. The usual redefinition of the scalar field follows (erroneously),

\beq \label{deffields2} \left({\partial \varphi \over \partial \phi } \right)^2 = {3 \over 4} \left( {\partial ~log~F(\phi) \over \partial \phi} \right)^2 + {Z(\phi) \over 2 F(\phi)}  ~~. \eeq

\noindent The usual redefinition \cite{analogy} is often presented as

\beq \left({\partial \varphi \over \partial \phi } \right)^2 = {3 \over 4} \left( {\partial ~log~F(\phi) \over \partial \phi} \right)^2 + {1 \over 2 F(\phi)} ~~,\eeq

\noindent which clearly corresponds to $Z(\phi) = 1$; this yields the canonical action

\beqa && S =  \int \sqrt{-g^*} ~d^4 x \times \\ \nonumber && \left[ R^*  -2 g^{*\mn} \varphi_{,\mu} \varphi_{,\nu}  - 4 U(\varphi)+ A^4(\varphi) {\cal L} (A^2(\varphi) g^*_\mn,\chi) \right]  ~~,\eeqa

\noindent which can be matched with action Eq. (\ref{actionZ}) through the relation

\beqa && -2 \varphi_{,\mu} \varphi_{,\nu} = -2 \left({\partial \varphi \over \partial \phi} \right)^2 \phi_{,\mu} \phi_{,\nu} = \\ \nonumber && -\left[  {3 \over 2} \left( {\partial ~log~F(\phi) \over \partial \phi} \right)^2  {1 \over  F(\phi)}  \right]  \phi_{,\mu} \phi_{,\nu}  ~~. \eeqa

However, the above shows that a proper treatment should use $Z(\phi) = 0$, as there is no intrinsic kinetic term in the original, Jordan frame theory Eq. (\ref{actionsinglescalar}).  This redefinition of the scalar field will affect the calculation of $\al = \partial ~log~A/ \partial \varphi$ and, as a consequence, yield incorrect predictions for the PPN parameters $\be$ and $\ga$; in particular, one obtains a dependence on $F(\phi)$ which would otherwise be missing. This can be seen from the following expressions \cite{parameters},

\beqa \label{parameters} \ga -1 &=& - {F'(\phi)^2 \over Z(\phi) F(\phi) + 2 F'(\phi)^2 }~~, \\ \nonumber \be -1 &=& {1 \over 4 } {F(\phi) F'(\phi) \over 2 Z(\phi) F(\phi) + 3 F'(\phi)^2} {d \ga \over d \phi}~~.\eeqa

\noindent which, for $Z(\phi) = 1$, yields the PPN parameters $\be$ and $\ga$ used in Refs. \cite{PPN}.

Hence, it appears that the PPN coefficients calculated for a wide variety of $f(R)$ models, and obtained by the dynamical identification $\phi = R$, are inaccurate: by reinstating the correct factor $Z(\phi) = 0$ into Eq. (\ref{deffields2}), one recovers

\beqa \left({\partial \varphi \over \partial \phi } \right)^2 &=& {3 \over 4} \left( {\partial ~log~F(\phi) \over \partial \phi} \right)^2 \rightarrow \\ \nonumber \varphi &=& {\sqrt{3} \over 2}\log~F(\phi)~~, \eeqa

\noindent so that the calculations for the PPN parameters $\ga = 1/2$ and $\be =1$ follow as argued previously -- and Eqs. (\ref{parameters}) clearly show. Moreover, notice that (as already discussed) one cannot simply identity GR with the limit $\ep \rightarrow 0 $ of a model with $f(R) = R + \ep \de_1(R)$, since  the corresponding limit of Eqs. (\ref{parameters})  (taking the correct factor $Z(\phi) = 0$) is ill-defined for $F'(\phi)=\ep \de''(\phi) \rightarrow 0$.

\begin{acknowledgments}

The authors would like to thank S. Capozziello for fruitful and elucidating discussions. The work of J.P. is sponsored by the FCT under the grant $BPD~23287/2005$. O.B. acknowledges the partial support of the FCT project $POCI/FIS/56093/2004$.

\end{acknowledgments}

\end{document}